\documentclass{article}

\usepackage{amsfonts, mathrsfs, amssymb, amsmath, enumerate, bbm, bm, mathtools}
\usepackage{multirow, multicol}
\usepackage{algorithm, algpseudocode} 
\usepackage{xcolor}
\usepackage{subcaption}

\def \longtitle {Weighted Sum-of-Trees Model for Clustered Data}
\def \shorttitle {\longtitle}

\definecolor{my_blue}{RGB}{65, 105, 225}

\usepackage{arxiv}
\usepackage[utf8]{inputenc} 
\usepackage[T1]{fontenc}    
\usepackage[]{hyperref}       
\usepackage{url}            
\usepackage{booktabs}       
\usepackage{nicefrac}       
\usepackage{microtype}      
\usepackage{cleveref}       
\usepackage{graphicx}
\usepackage{natbib}
\usepackage{doi}

\title{\longtitle}
\date{\today}

\author{
    \href{https://orcid.org/0000-0002-3570-6826}{%
        \raisebox{-0.1em}{\includegraphics[height=0.9em]{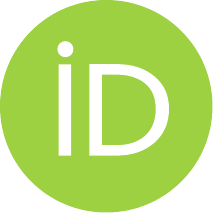}}%
        \hspace{1mm}Kevin McCoy%
    } \\
    Department of Statistics\\
    Rice University\\
    Houston, TX 77005 \\
    \texttt{kmm12@rice.edu} 
    \\[-1.5em] \hphantom{The University of Texas MD Anderson Cancer Center}
    \And
    \href{https://orcid.org/0000-0002-7827-1586}{%
        \raisebox{-0.1em}{\includegraphics[height=0.9em]{orcid.pdf}}%
        \hspace{1mm}Zachary Wooten%
    } \\
    Department of Biostatistics\\
    St. Jude Children's Research Hospital\\
    Memphis, TN 38105\\
    \texttt{zachary.wooten@stjude.org} 
    \\[-1.5em] \hphantom{The University of Texas MD Anderson Cancer Center}
    \AND \AND
    \href{https://orcid.org/0000-0003-0553-8056}{%
        \raisebox{-0.1em}{\includegraphics[height=0.9em]{orcid.pdf}}%
        \hspace{1mm}Katarzyna Tomczak%
    } \\
    Department of Translational Molecular Pathology\\
    The University of Texas MD Anderson Cancer Center\\
    Houston, TX 77030\\
    \texttt{kjtomczak@mdanderson.org} \\
    \And
    \href{https://orcid.org/0000-0003-3316-0468}{%
        \raisebox{-0.1em}{\includegraphics[height=0.9em]{orcid.pdf}}%
        \hspace{1mm}Christine B. Peterson\thanks{Author to whom correspondence should be addressed.}%
    } \\
    Department of Statistics\\
    Rice University\\
    Houston, TX 77005 \\
    \texttt{cbpeterson@rice.edu}
    \\[-1.5em] \hphantom{The University of Texas MD Anderson Cancer Center}
}

\hypersetup{
pdftitle={\longtitle},
pdfsubject={stat.ML},
pdfauthor={Kevin McCoy, Zachary Wooten, Katarzyna Tomczak, Christine B. Peterson},
pdfkeywords={decision trees, random forest, mixed effects modeling, weighted ensemble, clustered data},
}

\begin{document}
\maketitle

\maketitle

\begin{abstract}\label{sec:abstract}
Clustered data, which arise when observations are nested within groups, are incredibly common in clinical, education, and social science research. Traditionally, a linear mixed model, which includes random effects to account for within-group correlation, would be used to model the observed data and make new predictions on unseen data. Some work has been done to extend the mixed model approach beyond linear regression into more complex and non-parametric models, such as decision trees and random forests. However, existing methods are limited to using the global fixed effects for prediction on data from out-of-sample groups, effectively assuming that all clusters share a common outcome model. We propose a lightweight sum-of-trees model in which we learn a decision tree for each sample group. We combine the predictions from these trees using weights so that out-of-sample group predictions are more closely aligned with the most similar groups in the training data. This strategy also allows for inference on the similarity across groups in the outcome prediction model, as the unique tree structures and variable importances for each group can be directly compared. We show our model outperforms traditional decision trees and random forests in a variety of simulation settings. Finally, we showcase our method on real-world data from the sarcoma cohort of The Cancer Genome Atlas, where patient samples are grouped by sarcoma subtype.
\end{abstract}


\section{Introduction}\label{intro}



The decision tree is one of the most widely used machine learning methods. Its most popular implementation, the Classification and Regression Tree (CART) algorithm, was introduced by \citet{breiman_classification_1998}. Decision trees are built by recursively splitting the data space into rectangular partitions. Although decision trees work well in numerous scenarios, they can often become too large and start overfitting the training data. To avoid this, ensemble tree methods, such as random forests (RF) or boosting models, are used. A random forest fits trees in parallel using a bootstrapping technique to reduce overall model variance \citep{breiman_random_2001}, while boosting fits sequential trees to residuals in order to iteratively improve the model and reduce bias \citep{hastie_elements_2001}. Decision trees and their extensions are also tolerant of missing data, interpretable, robust against collinearity, and have built-in feature selection.

In clinical, education, and social science research, grouped data are incredibly common. For example, in educational research, students may be grouped within classrooms or schools. In biomedical research, patients may be grouped within hospitals or by disease subtype.
In these cases, the observations can no longer be considered independent due to the within-group, or intraclass, correlations \citep{jiang_linear_2021}. While independence among observations is not always a rigid assumption of decision trees and other machine learning models, accounting for the grouping structure during model construction can improve predictive performance. 
Handling of the grouped observations becomes especially important when predicting outcomes for observations that belong to groups outside of our training sample. For example, we may want to make predictions for students attending schools that were not included in our training dataset.

Clustered data are just one example of non-independent data structures, which are typically handled using mixed effects models. Mixed effects refer to the idea that some associations, known as fixed effects, share a consistent pattern across the population, while other attributes of the data have a random effect on the outcome, leading to correlation across multiple observations from a unit. For example, in a setting with repeated measures for individual patients, a mixed model might include a random effect per patient; for a setting where patients are clustered by hospital or clinician, those factors would be represented by random effects.
Some work has been done to incorporate mixed effects into tree-based models for clustered or longitudinal data \citep{hajjem_mixed_2011, spanbauer_nonparametric_2021}. However, existing methods suffer from limitations inherent in mixed effects models themselves. Mainly, out-of-sample prediction for mixed models uses only the fixed effects, as the random effects cannot be learned with no training data from those groups. Many methods also use a rigid additive structure for combining the fixed and random effects, similar to a linear mixed model, which takes away from the flexibility of decision trees. 

In Section \ref{sec:bkgd}, we discuss existing tree-based models for clustered data  and their limitations. 
In Section \ref{methods}, we introduce a lightweight method that learns a unique decision tree for each group. Predictions for out-of-sample groups are then calculated as a weighted linear combination of predictions from the decision trees learned from groups in the training data. In Section \ref{Simulated Results}, we highlight simulation studies that show that our method outperforms standard decision trees, random forests, and linear mixed models across a range of settings. Finally, in Section \ref{Real Results}, we show that our method performs competitively on a real data application to predict T-cell abundances for sarcoma subtypes not present in the training data.

\subsection{Background on Existing Methods} \label{sec:bkgd}


Generalized linear mixed models (GLMM) are the most standard approach for settings with clustered data.
In a GLMM, we fit the model $g(\mathbb{E}[y|X]) = X\beta + Z\alpha$, where $g(\cdot)$ represents the link function, $y$ corresponds to  the outcome variable, $X$ represents the design matrix for the fixed effects, and $Z$ is the design matrix for the random effects \citep{breslow_approximate_1993}.
As in standard linear models, GLMMs assume additive linear effects, which may be overly simplistic and lead to poor performance on complex datasets where there may be predictors with extreme values, non-linear effects, interactions, or other deviations from the assumptions of the linear model structure.


Mixed effects regression trees (MERT) \citep{hajjem_mixed_2011} use an expectation maximization (EM) approach to iteratively fit the model $y = f(X) + Z\alpha + \varepsilon$. The fixed effects $f(X)$ are fit using a regression tree, followed by an update fitting the random effects using standard techniques. MERT has been extended to forests using a bootstrapping technique \citep{hajjem_mixed-effects_2014}.  Currently, there are no extensions of MERT addressing categorical outcome data.

The MERT approach has a number of limitations. First, as in a standard GLMM model, the random part $Z\alpha$ is strictly linear and additive. MERT also assumes observations in different clusters are independent, though this is often not the case. Furthermore, the fixed and random effects are estimated separately, which means that any interaction  between fixed and random effects is precluded. Finally, as with standard mixed models, MERT can only use the fixed effects to make out-of-sample predictions. 

Case-specific random forests (CSRFs) \citep{xu_case-specific_2016} build a unique random forest for each new point in the test dataset. Instead of the traditional bootstrap technique, with a uniform selection probability for each observation, larger weights are assigned to points that are closer to the test case. One significant downside to this approach is the computational complexity and time needed to build a new random forest for each sample in the test dataset. This method also does not directly account for mixed effects, but nevertheless could be useful in settings with mixed effects present.

\citet{de_jong_developing_2021} propose internal-external cross-validation, which incorporates the clustered structure into drawing training vs.\ validation vs.\ test datasets, and picks the model that performs the best on the most out-of-group samples. However, only one output model is chosen.

In the Bayesian framework, Bayesian Additive Regression Trees (BART) were originally proposed by \citet{chipman_bart_2010}.
\citet{spanbauer_nonparametric_2021} proposed mixedBART as an extension to BART that can handle mixed effects. Similar to both GLMMs and MERT, mixedBART fits an additive model, where fixed effects and random effects are fit sequentially and separately via  BART and a Bayesian mixed model, respectively. Here, the Bayesian framework is quite convenient, as these two approaches can easily be integrated into alternating steps of the Markov Chain Monte Carlo (MCMC) algorithm. 

One limitation of BART is its handling of categorical variables. 
Normally, BART one-hot encodes categorical variables, replacing each unique category with its own binary indicator. Thus, each split in the base decision tree is associated only with one level of the categorical variable.
\citet{deshpande_flexbart_2025} recently proposed the flexBART model, which overcomes this limitation by allowing multiple levels of a categorical variable to be assigned to either side of the split rule. However, neither BART nor flexBART allow for levels of a categorical variable to be present in the test data but not the training data.

\section{Methods} \label{methods}

\subsection{Proposed Approach}
We now describe our proposed approach for flexible prediction on clustered data. In developing our method, we are particularly interested in the challenge of obtaining predictions for observations that do not belong to groups seen in the training data.

Let $\mathbf{X}$ represent an $n \times p$ matrix corresponding to $n$ observations on $p$ features, and let $Y$ be an $n \times 1$ column vector of outcomes. In our training dataset, each observation comes from one of $j= 1, 2, \ldots, J$ known groups or clusters, such that the observations within each cluster are correlated. We call the $n \times 1$ column vector of group assignments $C$. The testing dataset follows the same structure, except our groups are now indexed $m=1, 2, \ldots, M$. Here, we focus on settings where the groups in the test data fall outside groups encountered in the training dataset.

Our method proceeds in two stages. First, we construct a classification model $f: \mathbf{X} \to C$ on the training data. 
Using this approach, each test observation has a vector of prediction probabilities associated with belonging to each of the $J$ groups. We take the output probabilities of belonging to a certain class and call them weights  $\boldsymbol{w} = (w_1, \ldots, w_J)$. 
We interpret these weights as a similarity metric for measuring the closeness of a test data point $X_t$ to groups in the training data, considering only the $\mathbf{X}$ feature space. If there is more than one observation in an unseen group $m$, we average the similarity weights across the group and use that as $\boldsymbol{w}$.

The weights we learn are similar to the idea of propensity scores used in causal inference. However, instead of trying to predict the inclusion of an observation in the treatment group, we are predicting the ``closeness'' of new groups to groups seen in the training data. Our approach also shares some of the spirit of stacked ensemble learning \citep{van2007super}, with the key difference that the base models are learned for distinct groups in the input data and that the weights used to combine the base models are subject specific.

In the second stage of the model, we learn a unique decision tree for each training group $j$. The predicted outcome $y_t$ for a test data point $X_t$ is then calculated as a linear combination of these $J$ trees, each weighted by its corresponding value in $\boldsymbol{w}$.

\begin{figure}[!ht]
\centering
\includegraphics[width=\linewidth]{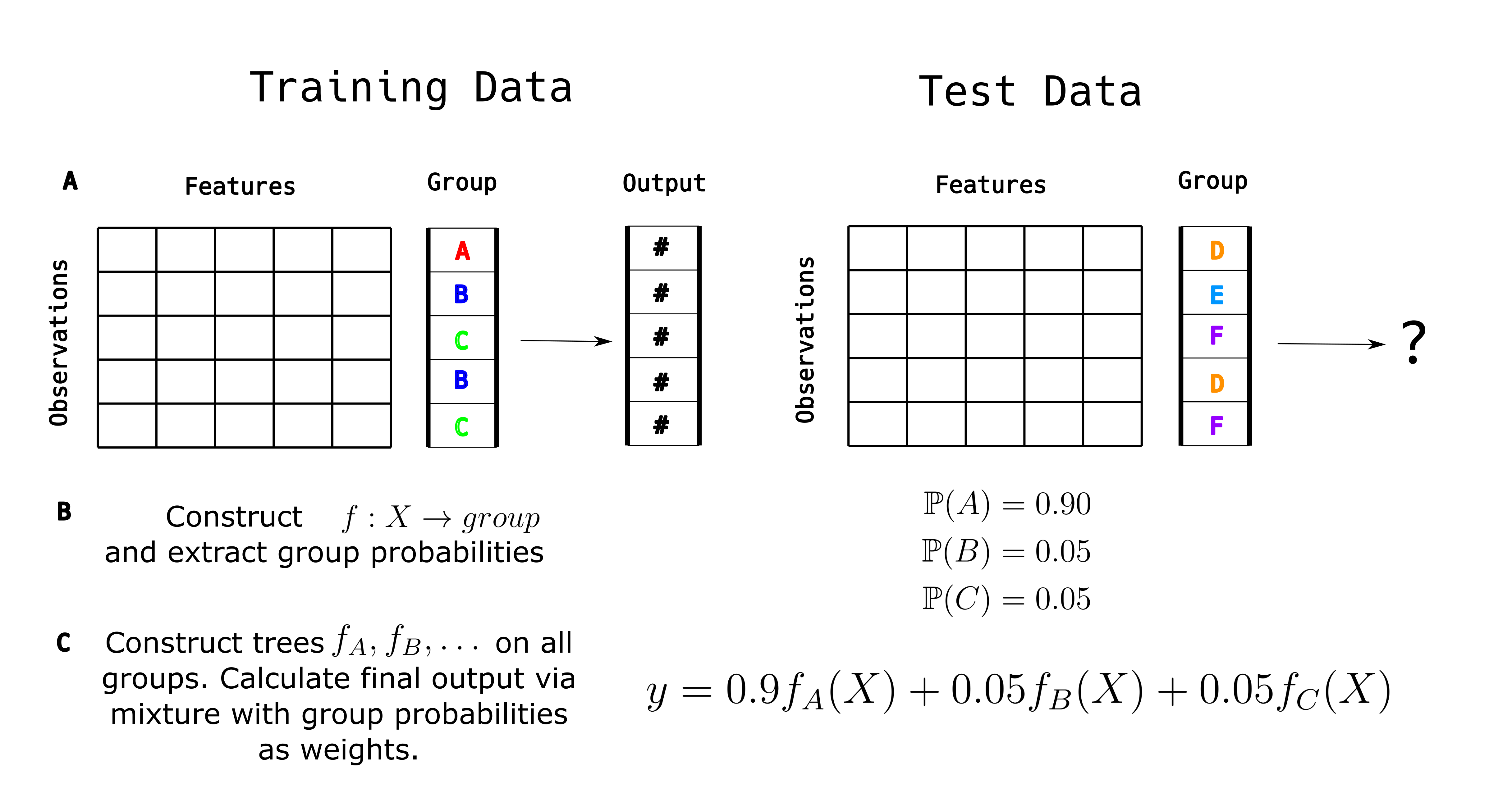} 
\caption{\textbf{(A)} Consider grouped data such that each training observation falls into some group (e.g., patients clustered by hospital or disease subtype). The test data are similarly structured but come from new groups not seen in the training data. \textbf{(B)} We then construct a classifier that predicts the group that each new observation belongs to, and we extract the predicted probabilities of group membership. \textbf{(C)} Finally, we construct independent trees or forests on each group seen in the training set. The final output is a linear combination of the predictions from all of these trees with the probabilities from (B) as weights.}
\label{fig:methods}
\end{figure}

For the first stage of the method, a model like logistic regression \citep{hosmer2013applied} or na{\"i}ve Bayes \citep{hastie_elements_2001} works well for its interpretable probabilities. In the second stage of the model, any learner can be used, but we focus on decision trees, as they strike a good balance between interpretability and predictive performance. Decision tress are often preferred for medical decision making, as they are easier for clinicians to understand \citep{banerjee2019tree}.   In our case study, we also consider random forests as the second-stage model, as they may offer more robust predictive performance.  \textbf{Figure \ref{fig:methods}} illustrates our proposed method.

\subsection{Implementation}
Our weighted sum-of-trees model is implemented in Python. For the base learners, we rely on the scikit-learn implementation of the methods with default parameter values \citep{pedregosa_scikit-learn_nodate}. The code is publicly available on GitHub at \href{https://github.com/kmccoy3/weighted-trees}{https://github.com/kmccoy3/weighted-trees}. 
 

\section{Results} \label{Results}
In this section, we compare the performance of our proposed weighted sum-of-trees method to a GLMM, a standard decision tree, and a random forest on both simulated and real data. We also consider a weighted combination of random forests, which is similar to our proposed method, but relies on a random forest instead of a decision tree in the second stage.  We are not able to compare our method to MERT, as there is no publicly available implementation of MERT. 


\subsection{Simulated Data} \label{Simulated Results}

\subsubsection{Data generation}

We construct a continuous response variable $y$ using the following model:
\begin{align*}
&y_{i,k} \sim \mathcal{N}(\mu(\boldsymbol{x}_{i,k}), 1) \\
&i \in \{1, \ldots, n_k\}, ~~~ k \in \{1, \ldots, K\},
\end{align*}

where $\mu$ is the mean function and the index $i$ denotes the $i$th subject within each group. The index $k$ indexes all groups, such that $K=J+M$.
We include 80\% of the groups $K$ in the training data and reserve the remaining 20\% as test data, so  $n_\text{test}$, the total number of observations in the test data, is $0.2 \times n \times K$. 

In order to generate $X$ data with structured correlation, we employ the matrix normal distribution:

$$\mathbf{X} \sim \mathcal{MN}_{K\times p} (\mathbf{M}, \mathbf{U}, \mathbf{V}),$$

where $\mathbf{M}$ is a $(K\times p)$ location matrix, $\mathbf{U}$ is the $(K \times K)$ group covariance matrix, and $\mathbf{V}$ is the $(p \times p)$ feature covariance matrix. The matrix normal distribution allows us the freedom to specify a number of correlation structures between not only features, but also groups in settings with clustered data. We generate $n$ samples from this distribution, resulting in a total of $n \times K$ observations.

We consider three distinct simulation settings, each of which will have its own mean function $\mu$. The first considers nonlinear fixed effects with independent linear random effects, the second  considers nonlinear fixed effects with dependent linear random effects, and the third considers the case where the underlying function generating the data varies between groups. We discuss each setting below.

\textit{Simulation setting 1:}
In this simulation, we set 
$$\mu_0 = f(\boldsymbol{x}_{i,k}) + \mathbf{Z}\boldsymbol{\alpha}_k$$
where $f$ is the fixed effects function, $\mathbf{Z}$ is the random effects design matrix, and $\boldsymbol{\alpha}_k$ are the random slopes and intercept shared by all observations within the $k$th group. We employ Friedman's five-dimensional test function \citep{friedman_multivariate_1991} for the fixed effects function $f$:

$$f(\boldsymbol{x}_{i,k}) = \sin(\pi x_1 x_2) + 2(x_3-0.5)^2 + x_4 + 0.5 x_5.$$

This function is a popular benchmark for machine-learning methods, as it includes a mix of linear, non-linear, and interaction terms. We use a $\text{Unif}(-1,1)$ distribution to generate the entries of $\mathbf{M}$, with $\mathbf{U}$ and $\mathbf{V}$ being identity matrices. We then generate the random effects $\boldsymbol{\alpha}_k$  using the same matrix normal distribution with the same row covariance matrix $\mathbf{U}$. Taken together, this is equivalent to generating one sample from a matrix normal distribution $\mathcal{MN}_{K\times (p+1)} (\mathbf{0}, \mathbf{U}, \mathbf{V}^*)$, where $\mathbf{V}^*$ is the column covariance matrix specifying the variance structure between the random slopes and intercept. We use $\mathbf{V}^*=\sigma_\alpha^2 \times \mathbf{I}$, where $\mathbf{I}$ is the $(p+1) \times (p+1)$ identity matrix. This data generation strategy results in similar random effects for groups that are close in the $X$ space.

\textit{Simulation setting 2:}
Next, we conducted a similar experiment, where instead of using $\mathbf{U}=\mathbf{I}$, we generated a unique $\mathbf{U}$ matrix from an inverse-Wishart distribution $\mathcal{W}^{-1}(\mathbf{I}, K+1)$, where $K+1$ is the degrees of freedom and the scale matrix is taken as the $K \times K$ identity matrix $\mathbf{I}$. We do this to simulate settings where the feature data and the random effects have a similar group covariance structure.

\textit{Simulation setting 3:}
In this scenario, we consider multiple data generating processes  similar to those in \citep{deshpande_flexbart_2025}. Here, $X$  is generated identically to \textit{Simulation setting 1}, while the entries of $\mathbf{M}$ are now generated from a $\text{Unif}(0,2)$ distribution. This simulation study was designed to test scenarios where the data is governed by multiple basis functions, and each group in the data is governed by some, but not all, of the basis functions. Similar to \citet{deshpande_flexbart_2025}, we also generate 5 extra noise variables so that $p=10$. 
We have the following basis functions:
\begin{align*}
     f_0(x) &= 10\sin(\pi x_1x_2) \\
    f_1(x) &= 10(x_3-0.5)^2 \\
    f_2(x) &= 10(x_1 - 0.5)^2 + 10x_2 + 5x_3 \\
     f_3(x) &= 6x_1 + (4-10*\mathbbm{1}(x_2 > 0.5)) * \sin(\pi x_1) - 4*\mathbbm{1}(x_2 > 0.5) + 15
\end{align*}
These functions are then combined in the following data generating processes:
\begin{align*}
    \mu_1(x, k) &= \left(f_0(x) + f_1(x) + f_2(x) - 0.75\right) \cdot \mathbbm{1}\{k \in C_0, C_2, C_4, \dots\} \, +\\ &\qquad \qquad f_3(x) \cdot \mathbbm{1}\{k \in C_1, C_3, C_5, \dots\} \\
    \mu_2(x, k) &= f_0(x) \cdot \mathbbm{1}\{k \in C_0, C_3, C_6, \dots\} + f_1(x) \cdot \mathbbm{1}\{k \in C_1, C_4, C_7, \dots\} \, + \\ &\qquad \qquad  f_2(x) \cdot \mathbbm{1}\{k \in C_2, C_5, C_8, \dots\} + f_3(x) \cdot \mathbbm{1}\{k \in C_0, C_2, C_4, \dots\}\\
    \mu_3(x, k) &= f_0(x) \cdot \mathbbm{1}\{k \in C_0, C_1, C_3, C_4, \dots\} + f_1(x) \cdot \mathbbm{1}\{k \in C_1, C_2, C_4, C_5, \dots\} \, + \\ &\qquad \qquad f_2(x) \cdot \mathbbm{1}\{k \in C_0, C_2, C_3, C_5, \dots\} + f_3(x) \cdot \mathbbm{1}\{k \in C_0, C_2, C_4, \dots\}
\end{align*}

To apply  the competing methods, we relied on the Python packages sklearn and statsmodels. We use default parameter values for each method, except for random forests, which are implemented using $J$ trees since our weighted sum-of-trees method is also trained on $J$ trees. The weighted combination of random forests uses $J$ forests with $J$ trees each.

\subsubsection{Comparative performance}

\begin{figure}[!ht]
\centering
\includegraphics[width=\textwidth]{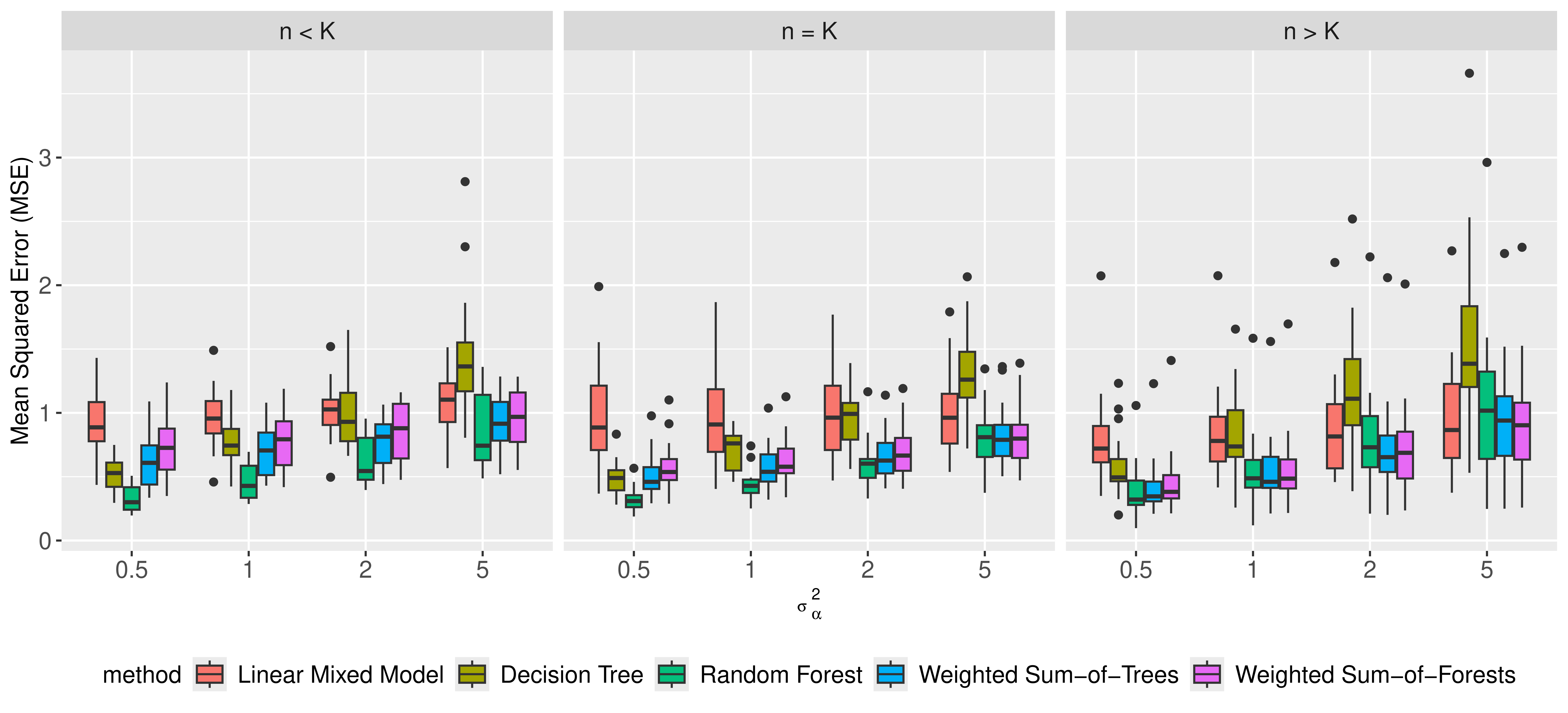}
\caption{\textit{Simulation setting 1}. Mean squared error (MSE) over a range of noise values $\sigma_\alpha$ for settings where $\mathbf{U} = \mathbf{I}$. The scale of the response variable $y$ is standardized. Each boxplot represents MSE values across 20 simulated datasets. Simulations were performed for the settings $n=10$, $K=40$ (left), $n=K=20$ (center), and $n=40$, $K=10$ (right). }
\label{fig:simulated-results-1}
\end{figure}

\textbf{Figure \ref{fig:simulated-results-1}} showcases the mean squared error (MSE) for \textit{Simulation setting 1}. The MSE is calculated on test data as $\frac{1}{n_\text{test}}\sum_{i=1}^{n_\text{test}} ( y_i - \hat{y}_i)^2$, where $\hat{y}_i$ is the predicted outcome value for observation $i$.  We considered  simulation settings with varying  number of groups $K$ and number of observations per group $n$. We also varied $\sigma^2_{\alpha}$, which scales the column covariance matrix for the random effects, so larger values of $\sigma^2_\alpha$ result in random effects with larger cross-group differences. 

Based on these simulation results, we find that our weighted sum-of-trees method outperforms standard decision trees and becomes competitive with random forest as the magnitude of random effects $\sigma^2_\alpha$ becomes larger. This is impressive given that our method does not incorporate bootstrapping of data or random selection of features. It is also fitted with only $J=0.8\times K$ trees. Interestingly, the decision tree's performance degrades as the amount of random noise is increased, while the linear mixed model's performance remains more stable across $\sigma^2_\alpha$ values. We also find that using random forests in the second stage of the model does not significantly change the performance of the overall method.

\begin{figure}[!ht]
\centering
\includegraphics[width=\textwidth]{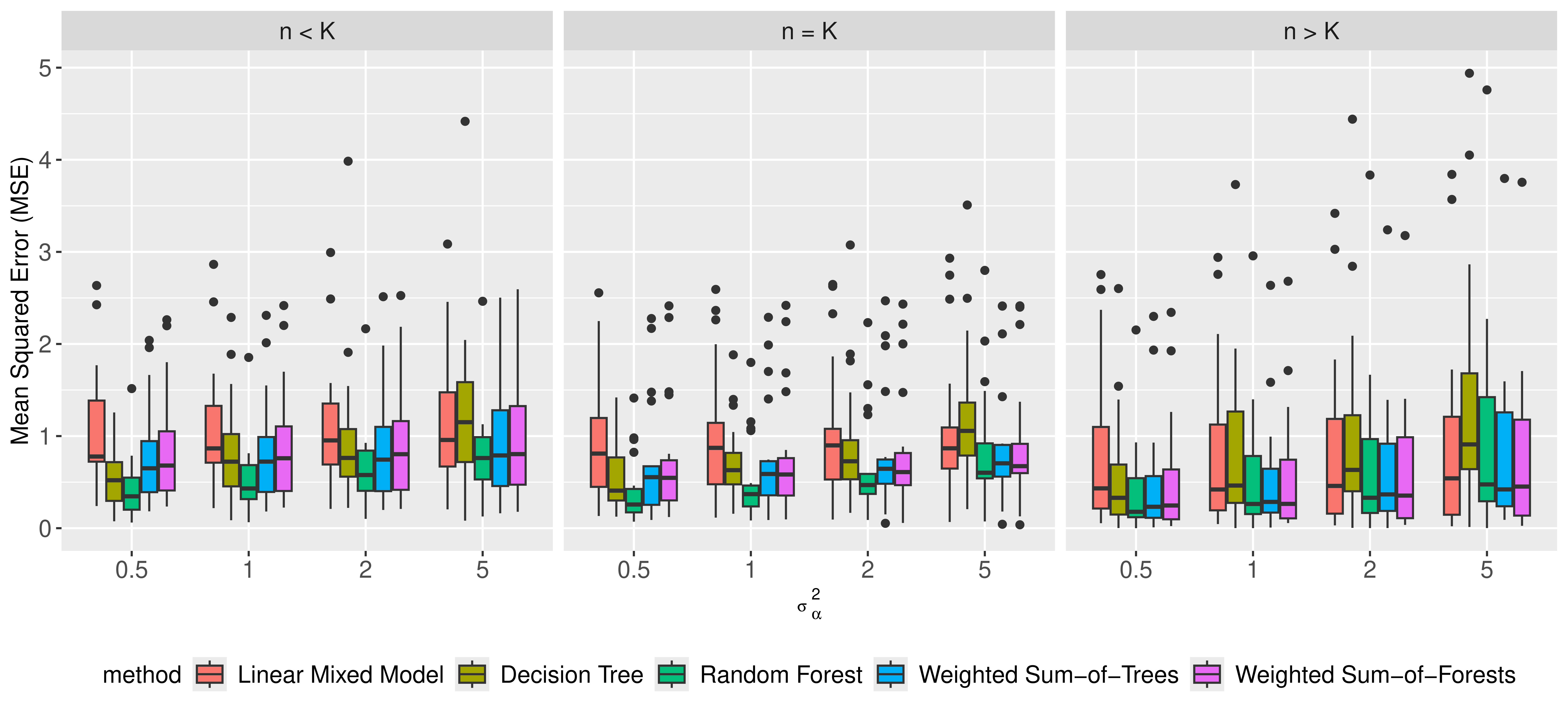}
\caption{\textit{Simulation setting 2}. Mean squared error (MSE) of the various methods tested over a range of random noise values $\sigma_\alpha$ for settings where $\mathbf{U}$ is generated from an inverse-Wishart distribution. The scale of the response variable $y$ is standardized. Each boxplot represents MSE values across 20 simulated datasets. Simulations were performed for the settings $n=10$, $K=40$ (left), $n=K=20$ (center), and $n=40$, $K=10$ (right).  }
\label{fig:simulated-results-2}
\end{figure}

 As shown in  \textbf{Figure \ref{fig:simulated-results-2}}, the results from \textit{Simulation setting 2} are much more homogeneous than in the setting with $\mathbf{U}=\mathbf{I}$, though the trends are generally similar. Linear mixed models perform adequately over the range of $\sigma_\alpha$ values, while decision trees and random forests, which do comparatively well for the settings with smaller $\sigma^2_\alpha$, exhibit progressively worse performance for higher values. Our weighted sum-of-trees method becomes more advantageous relative to traditional decision trees as the random effects grow larger, eventually becoming competitive with random forests. Notably, the scale of \textbf{Figure \ref{fig:simulated-results-2}} is larger than that of \textbf{Figure \ref{fig:simulated-results-1}}, as each method performed poorly for some of the simulated datasets.

\begin{figure}[!ht]
\centering
\includegraphics[width=\textwidth]{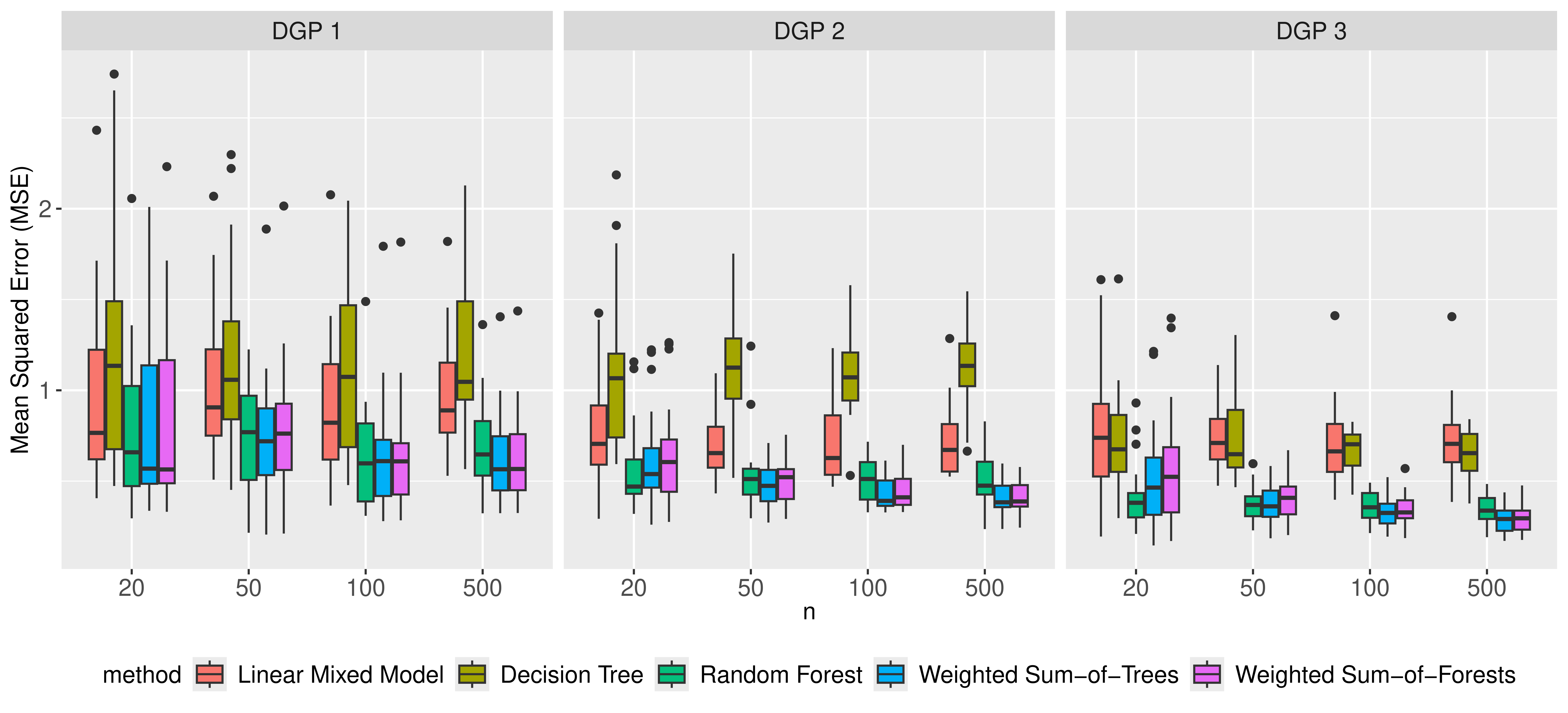}
\caption{\textit{Simulation setting 3}. Mean squared error (MSE) of the various methods tested over a range of values for the number of observations per group, $n$, for settings where data is generated according to one of 3 distinct data generating processes. Throughout all simulations, $K=20$ groups were used. The scale of the response variable $y$ is standardized. Each boxplot represents MSE values across 20 simulated datasets.
}
\label{fig:simulated-results-3}
\end{figure}

\begin{figure}[!ht]
    \centering
    \includegraphics[width=\linewidth]{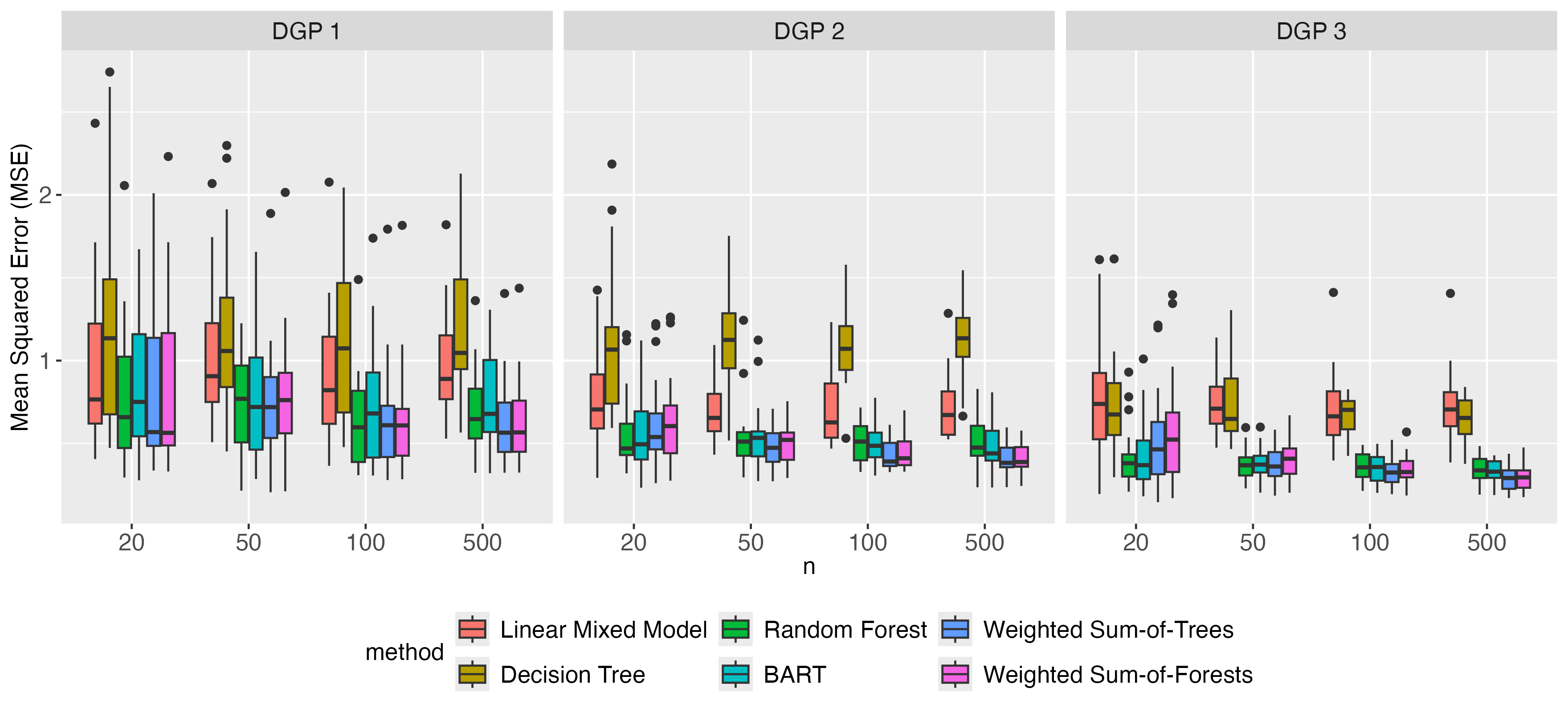}
    \caption{BART comparison}
    \label{fig:bart_comp}
\end{figure}

Finally, we consider \textit{Simulation setting 3}. The results are presented in \textbf{Figure \ref{fig:simulated-results-3}}. For all three data generating processes, the sum-of-trees method, sum-of-forests, and random forest methods perform better than linear mixed models and decision trees. Our sum-of-trees method overtakes random forest in mean performance as $n$ increases, though the distributions of MSE values do still overlap considerably. Notably, using random forests as the second-stage model does not considerably affect the performance of our method when compared to using decision trees. For this simulation setting, we also considered BART as a comparator; its performance was similar to that of random forests (\textbf{Figure \ref{fig:bart_comp}}).

We report the run times of these methods in \textbf{Table \ref{tab:runtime_data}}. Our weighted sum-of-trees method runs in approximately the same amount of time as a random forest for small $n$, which makes sense given that they incorporate the same number of trees. However, as $n$ grows larger, our method becomes more advantageous and runs in times more similar to base decision trees, owing to fewer observations being used to train each base tree. It is also worth noting that the first stage of our model (which in this case used linear regression) does not add any considerable overhead to the run time. Decision trees exhibited the fastest run times, while BART and linear mixed models were the slowest.

\subsection{Real Data Application} \label{Real Results}

We apply our proposed weighted sum-of-trees method to real data from The Cancer Genome Atlas (TCGA), where patients are naturally grouped by cancer subtype.

\subsubsection{Sarcoma data}

Sarcomas are a heterogeneous family of cancers arising from the connective tissues in the body, including bone, nerve, cartilage, muscle, fat, and vasculature. 
Since sarcomas are rare (less than 1\% of adult cancers), sarcoma patients have historically been grouped together in clinical trials. However, there is a strong interest in identifying treatments that may benefit specific subgroups \citep{katz_more_2018}. In particular, immunotherapy drugs, which have revolutionized treatment for many cancer types but met with mixed results in sarcoma, may benefit patients with immune microenvironments conducive to therapeutic response \citep{ayodele_immunotherapy_2020}.

Here, we analyzed the data for the sarcoma cohort in TCGA \citep{lazar_comprehensive_2017}.
We attempt to predict the relative abundance of T-cells, a key marker associated with response to immunotherapy \citep{thorsson_immune_2018}. Our method is well suited for this application, as some types of sarcomas have very few observations in the TCGA dataset. We can improve predictive accuracy for these less common subtypes by aligning the predictive model for these smaller groups with those of the larger groups with more observations. Moreover, the TCGA dataset encompasses only 6 of the most common subtypes of sarcoma, out of more than 100 \citep{grunewald_sarcoma_2020}. Our approach could be used to generate predictions for subjects with rare disease subtypes that are not well characterized in existing datasets. The application of this model could therefore answer a key question in precision medicine for rare disease types: given a new patient's genomic profile for a subtype that has not been well represented in clinical trials, how likely are they to benefit from immunotherapy?

In the TCGA data, there are samples available for the following sarcoma subtypes: 80 leiomyosarcoma (LMS), 50 dedifferentiated liposarcoma (DDLPS), 44 undifferentiated pleomorphic sarcoma (UPS), 17 myxofibrosarcoma (MFS), 10 synovial sarcoma (SS), and 5 malignant peripheral nerve sheath tumor (MPNST) samples. 
Of the 80 LMS patients, there are 53 soft tissue LMS (STLMS) and 27 uterine LMS (ULMS). 
We use the samples from LMS, DDLPS, UPS, and MFS sarcoma types as training data and the samples from SS and MPNST as test data. 

For input features, we considered a combination of clinical and genomic variables likely to play a role in immune activity and tumor immune infiltration. From the clinical data, we include the following 
demographic variables: age at diagnosis and gender. We also include  copy number aberration of the following genes identified as frequently aberrant in the TCGA sarcoma dataset \citep{lazar_comprehensive_2017}: \textit{JUN}, \textit{VGLL3}, \textit{TERT}, \textit{MAP3K5}, \textit{UST}, \textit{CDKN2A}, \textit{YAP1}, \textit{CDKN1B}, \textit{PTPRQ}, \textit{RB1}, \textit{TP53}, \textit{MYOCD}, \textit{NF1}, \textit{CCNE1}, \textit{CEBPA}, \textit{ZNF552}, \textit{ATRX}, \textit{PTEN}, \textit{DDIT3}, \textit{CDK4}, \textit{HMGA2}, \textit{MDM2}, and \textit{FRS2}.
We consider the number of silent and non-silent mutations per megabase (Mb); an increased number of mutations is generally associated with increased neoantigen presentation and higher immune infiltration \citep{wang2021beyond}. Finally, we also include mRNA gene expression levels of the following genes, which have been previously identified as relevant to immunotherapy response \citep{dulal2023tackling}: \textit{CTLA4}, \textit{TIM3} (\textit{HAVCR2}), \textit{LAG3}, \textit{PD1} (\textit{PDCD1}), \textit{TCF7}, and \textit{TIGIT}. We use these features to predict the relative abundance of T-cells in the tumor, as a surrogate for immunotherapy response.

\begin{figure}[!ht]
\centering
\includegraphics[width=0.5\linewidth]{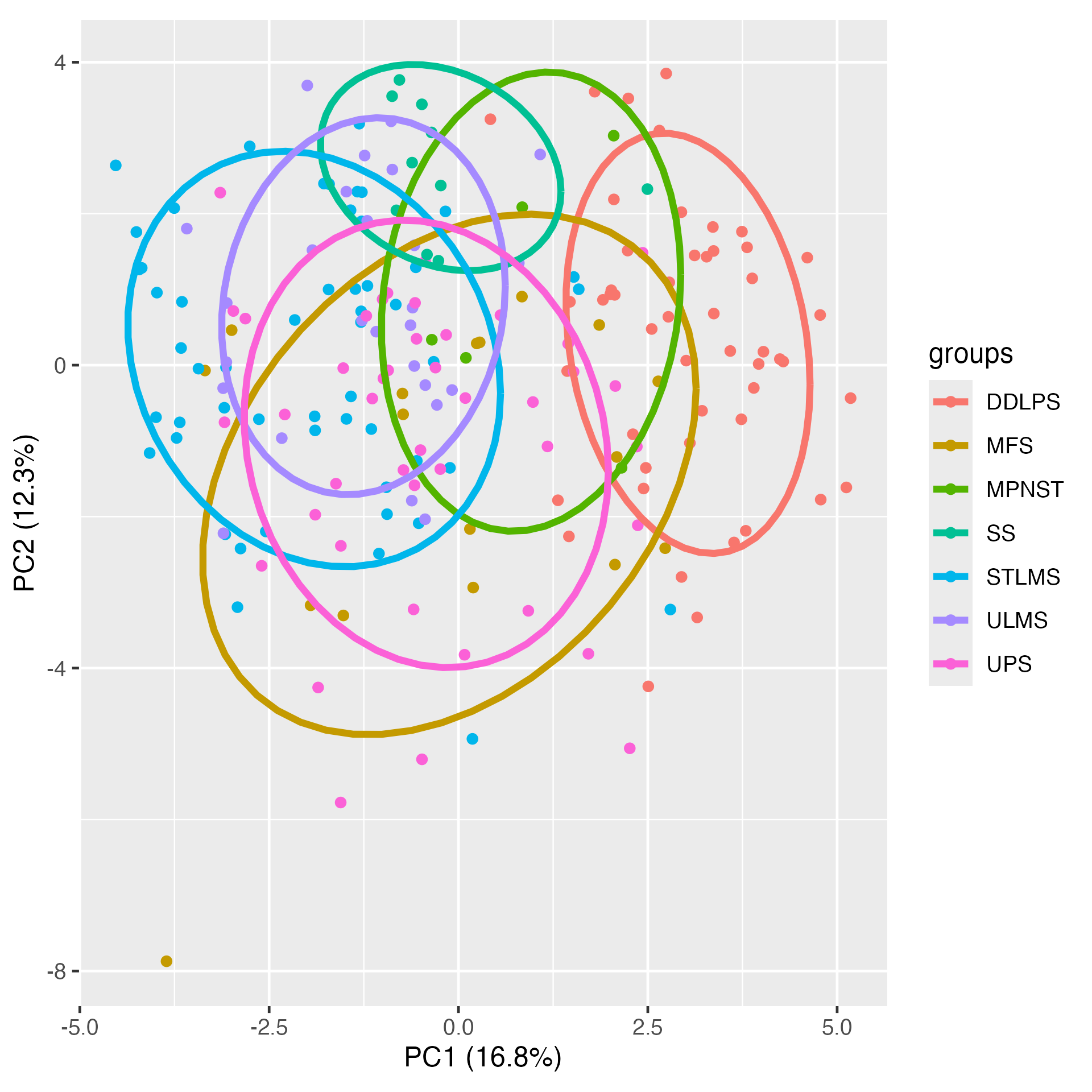} 
\caption{Principal component analysis of the The Cancer Genome Atlas (TCGA) dataset. The points are colored according to that observation's sarcoma subtype. Additionally, the ellipses drawn correspond to the normal data curve fit to each sarcoma subtype with coverage probability of 0.68. Key: DDLPS = dedifferentiated liposarcoma, MFS = myxofibrosarcoma, MPNST = malignant peripheral nerve sheath tumor, SS = synovial sarcoma, STLMS = soft tissue leiomyosarcoma, ULMS = uterine leiomyosarcoma, UPS = undifferentiated pleomorphic sarcoma}
\label{fig:pca}
\end{figure}

In order to visualize the data, we conduct principal component analysis on the input features and plot the top 2 principal components in \textbf{Figure \ref{fig:pca}}. Together, the first two principal components account for $29.1\%$ of the variance of the original dataset. Although there is a fair amount of heterogeneity within subtypes, distinct patterns are evident across the sample groups.

We compare our sum-of-trees method against a traditional LMM, a standard decision tree, and a random forest with 5 trees. We choose 5 trees because that is the same number of trees as in our sum-of-trees model. Finally, we test our method using random forests as the base learner instead of a decision tree and compare those results with a standard random forest with 100 trees. The random forests in our method use 20 trees, so in total 100 trees are also used. 
We find that our method with both decision trees and random forests as the base learner vastly outperform the LMM and basic decision tree in terms of achieving a lower MSE on the test set as a whole (\textbf{Table \ref{tab:real_data_performance}}). We also see that when the amount of trees is the same, our method does better than standard random forests. 

In \textbf{Table \ref{tab:learned_weights}}, we show the weights learned during the first stage of the model. These weights are generally concordant with the patterns shown in \textbf{Figure \ref{fig:pca}}, where SS is partially overlapping with ULMS, while MPNST is more heterogeneous, despite its small sample size.

\begin{figure}[!ht]
    \centering
    \begin{subfigure}[b]{0.32\textwidth}
        \includegraphics[width=\textwidth]{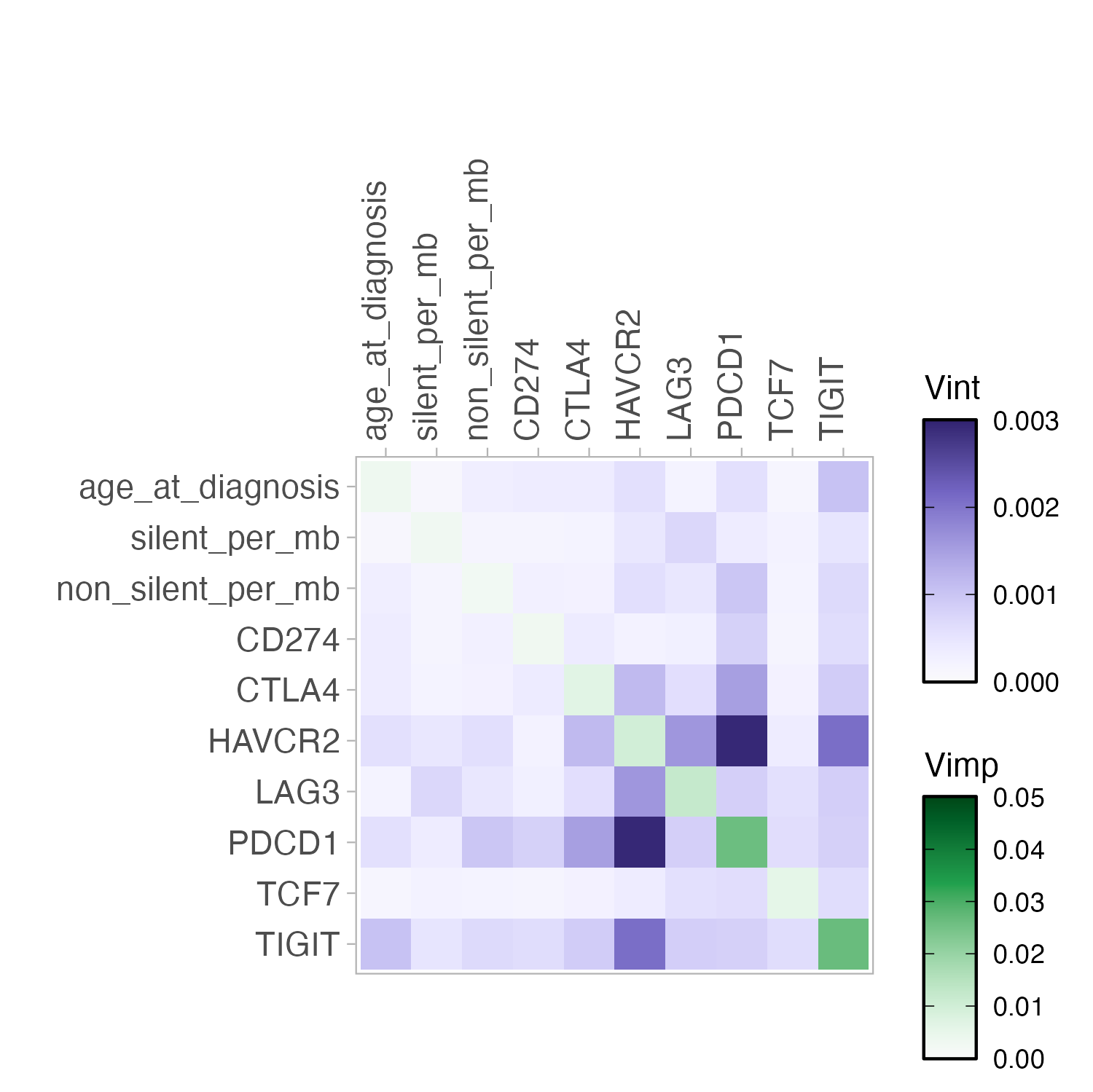}
        \caption{STLMS}
    \end{subfigure}
    \begin{subfigure}[b]{0.32\textwidth}
        \includegraphics[width=\textwidth]{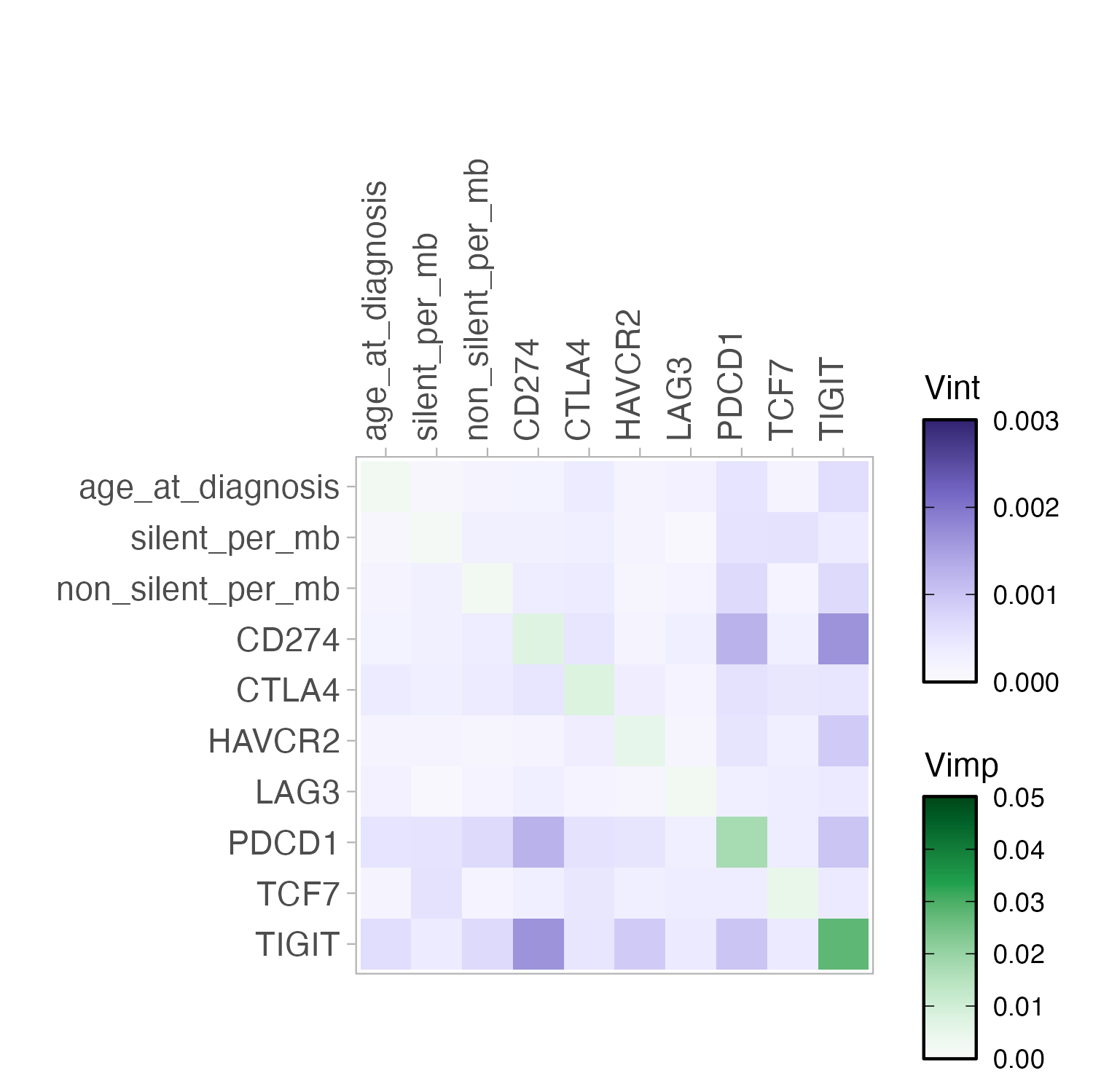}
        \caption{ULMS}
    \end{subfigure}
    \begin{subfigure}[b]{0.32\textwidth}
        \includegraphics[width=\textwidth]{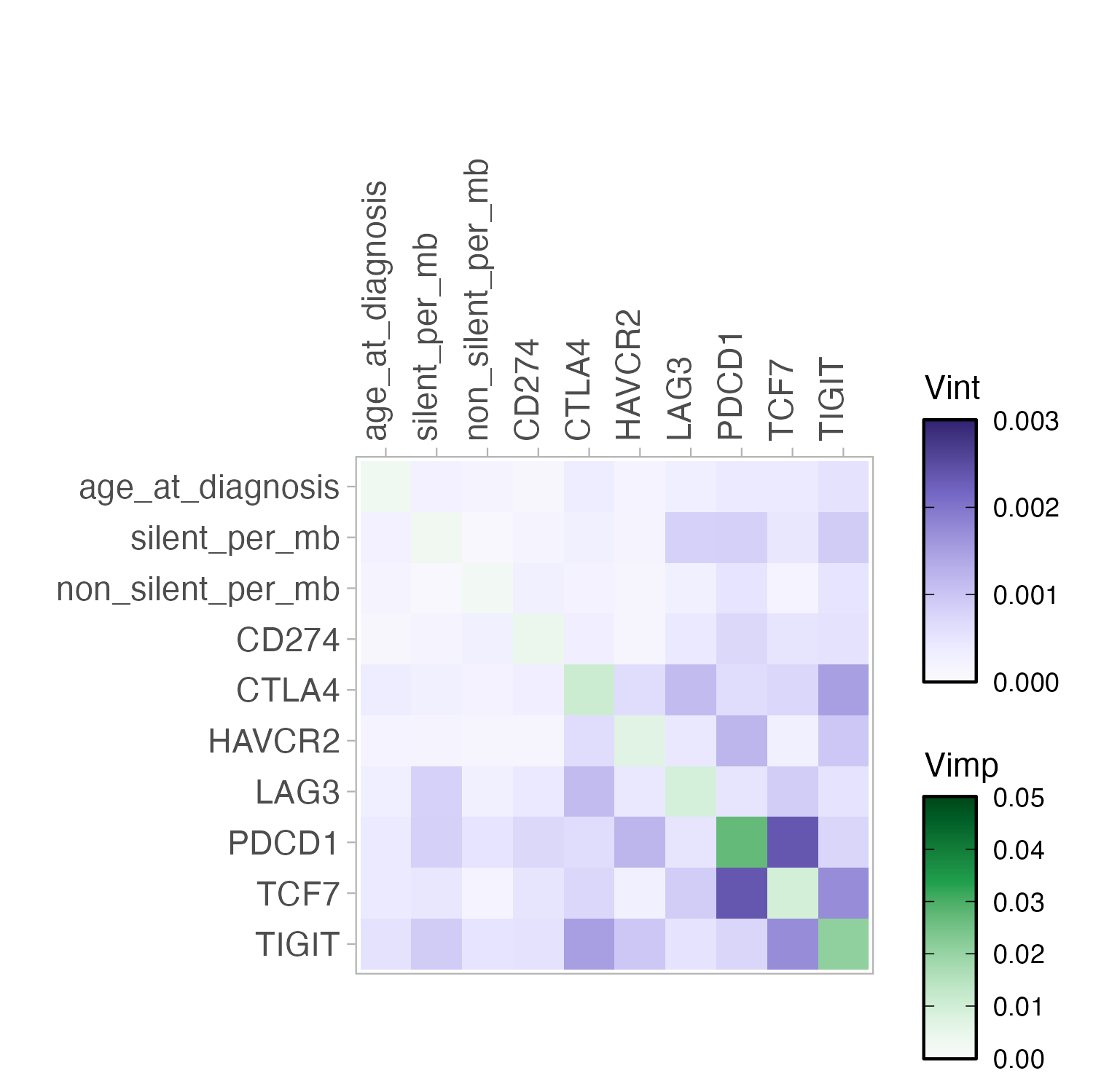}
        \caption{DDLPS}
    \end{subfigure}
    \\
    \begin{subfigure}[b]{0.32\textwidth}
        \includegraphics[width=\textwidth]{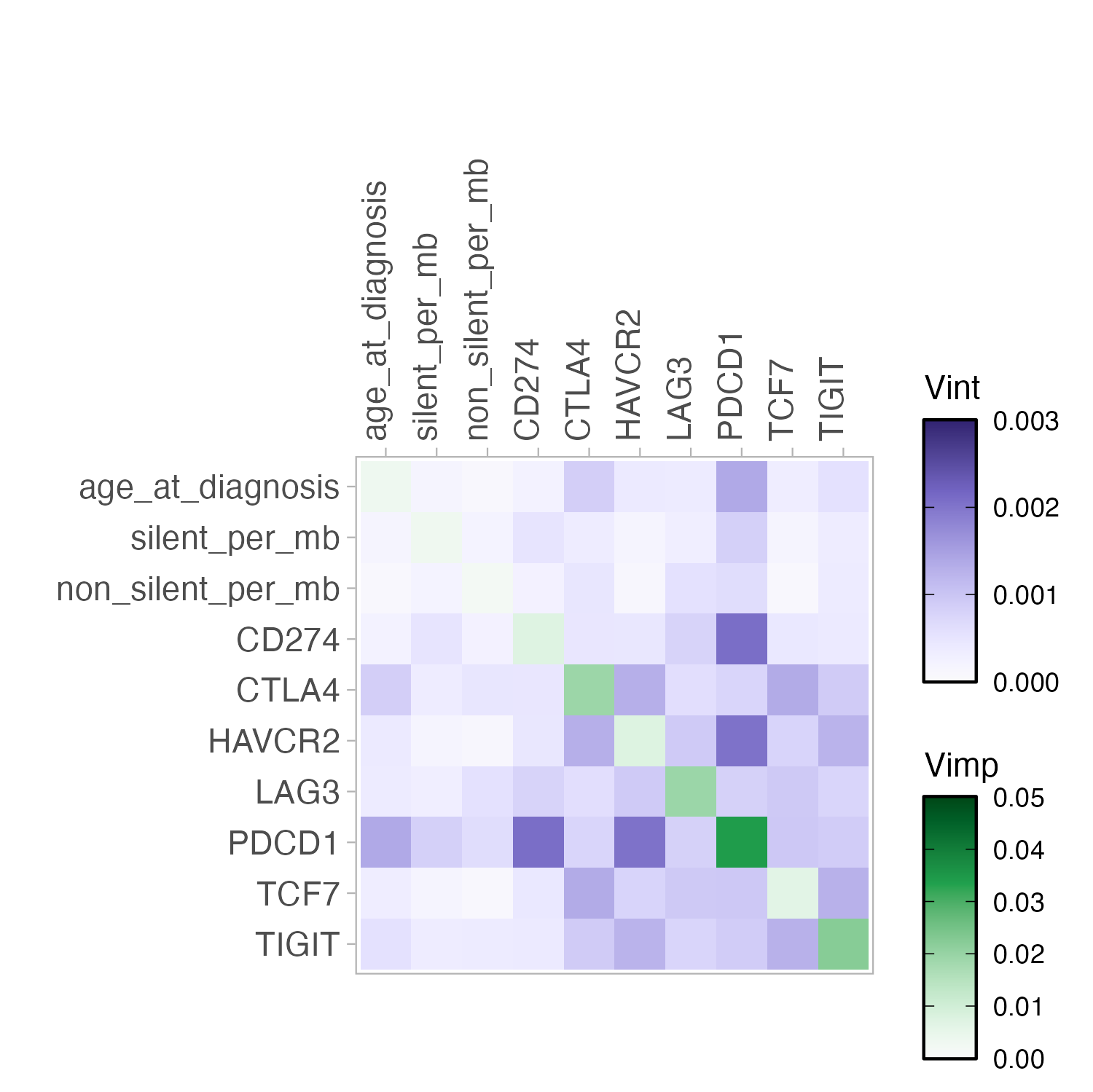}
        \caption{UPS}
    \end{subfigure}
    \begin{subfigure}[b]{0.32\textwidth}
        \includegraphics[width=\textwidth]{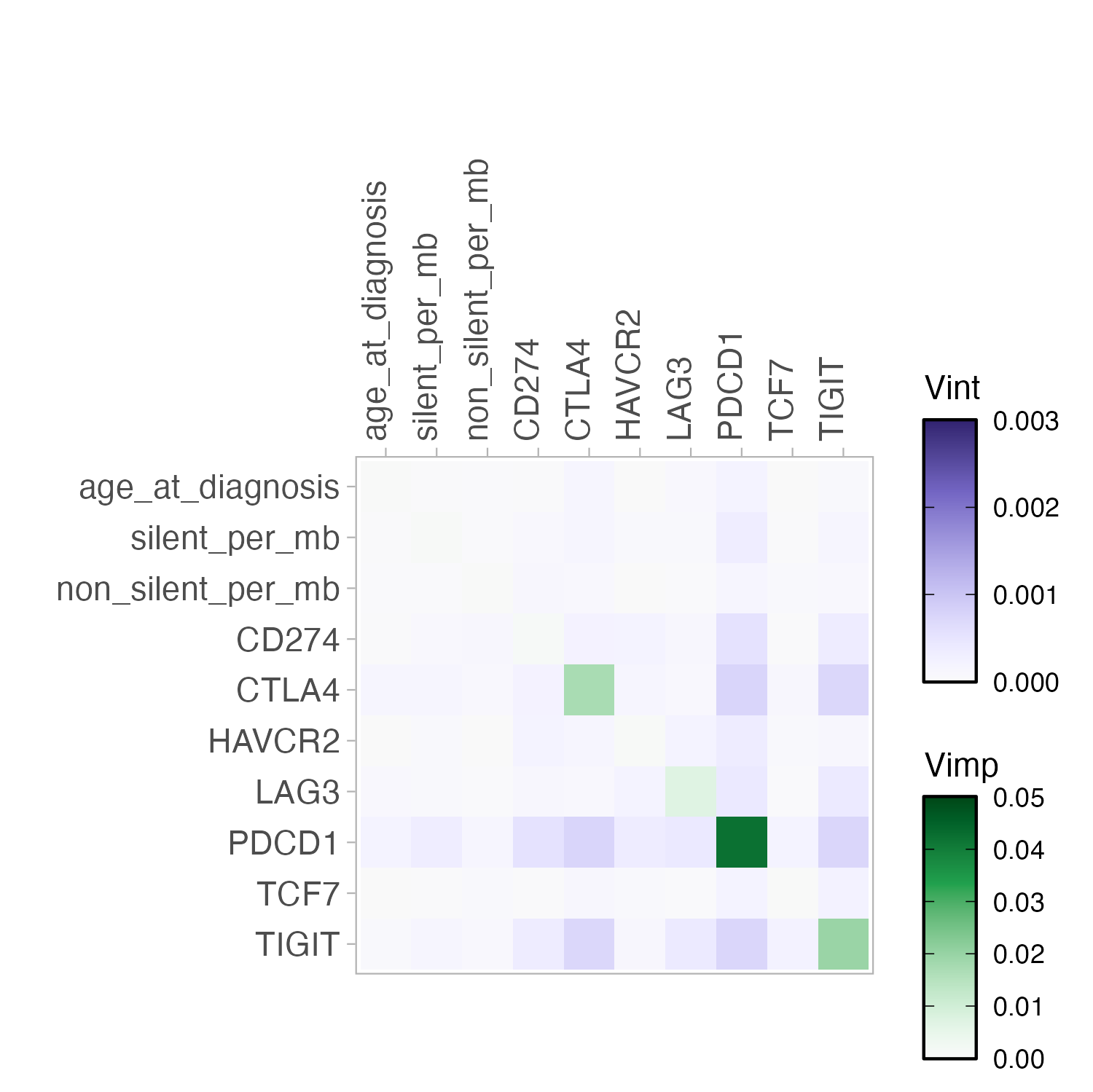}
        \caption{MFS}
    \end{subfigure}
    \caption{Variable importance (Vimp) and variable interaction (Vint) plots for a random forest trained on each specific sarcoma subtype. The variables of gender and the copy number aberrations were excluded as they were found to be much less importance and showed little interaction effects. All plots share the same scale.}
    \label{fig:vivi_plots}
\end{figure}

To show the flexibility and interpretability of our approach, we build a random forest on each sarcoma subtype. Then, in \textbf{Figure \ref{fig:vivi_plots}}, we present the group-specific variable importance and variable interaction (VIVI) plots \citep{inglis_vivid_2022} for each of the random forests learned. 
Although various copy number aberrations were informative in the first-stage model, we found that gender and the copy number aberration features had relatively smaller variable importances in the outcome model, and they did not show strong interaction effects. For this reason, we leave out these variables in \textbf{Figure \ref{fig:vivi_plots}}.

\begin{figure}[!ht]
    \centering
    \subcaptionbox{Synovial sarcoma (SS)}{
        \includegraphics[width=0.45\textwidth]{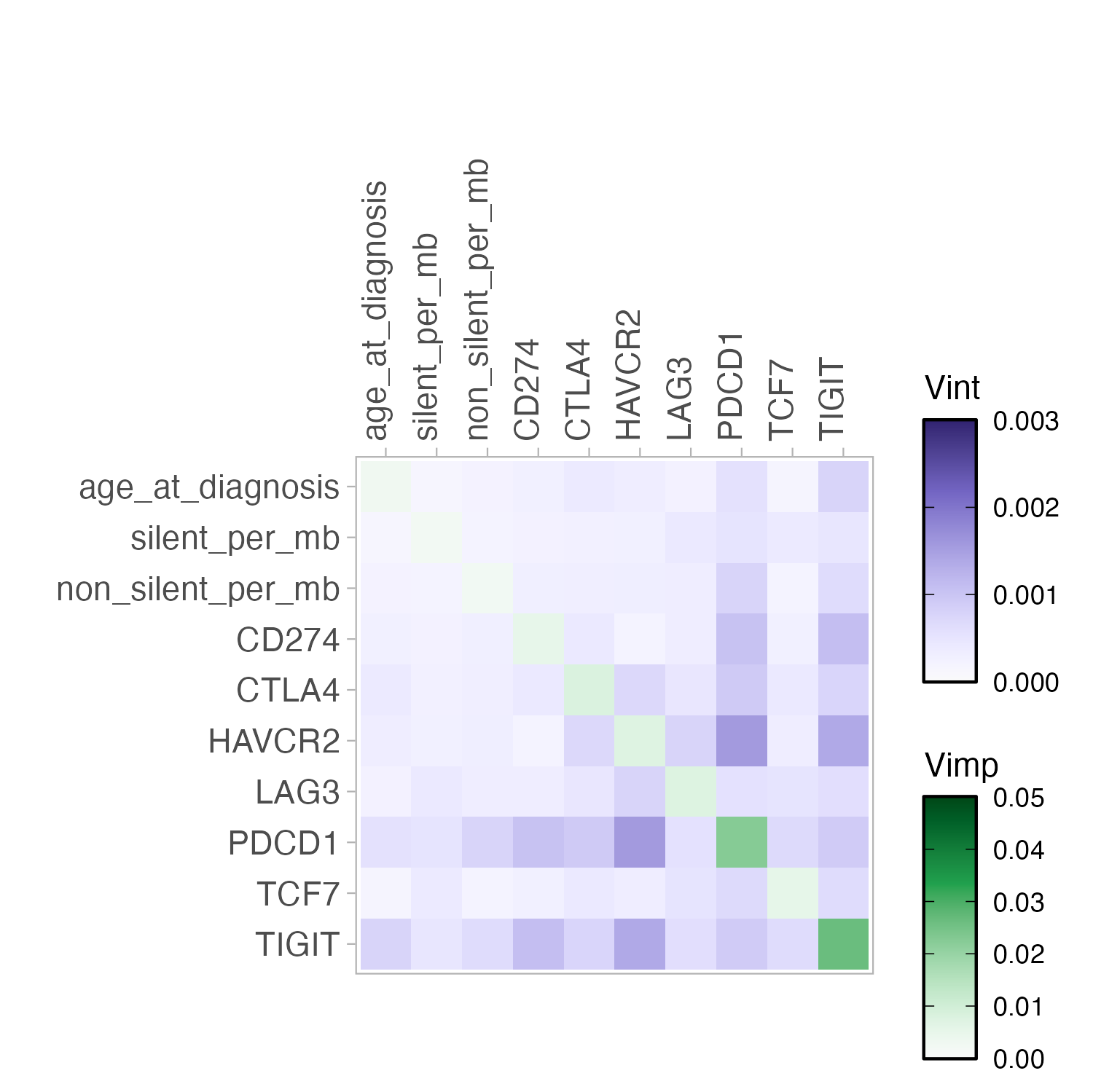}}  
    \hspace{1em}  
    \subcaptionbox{Malignant peripheral nerve sheath tumor (MPNST)}{
        \includegraphics[width=0.45\textwidth]{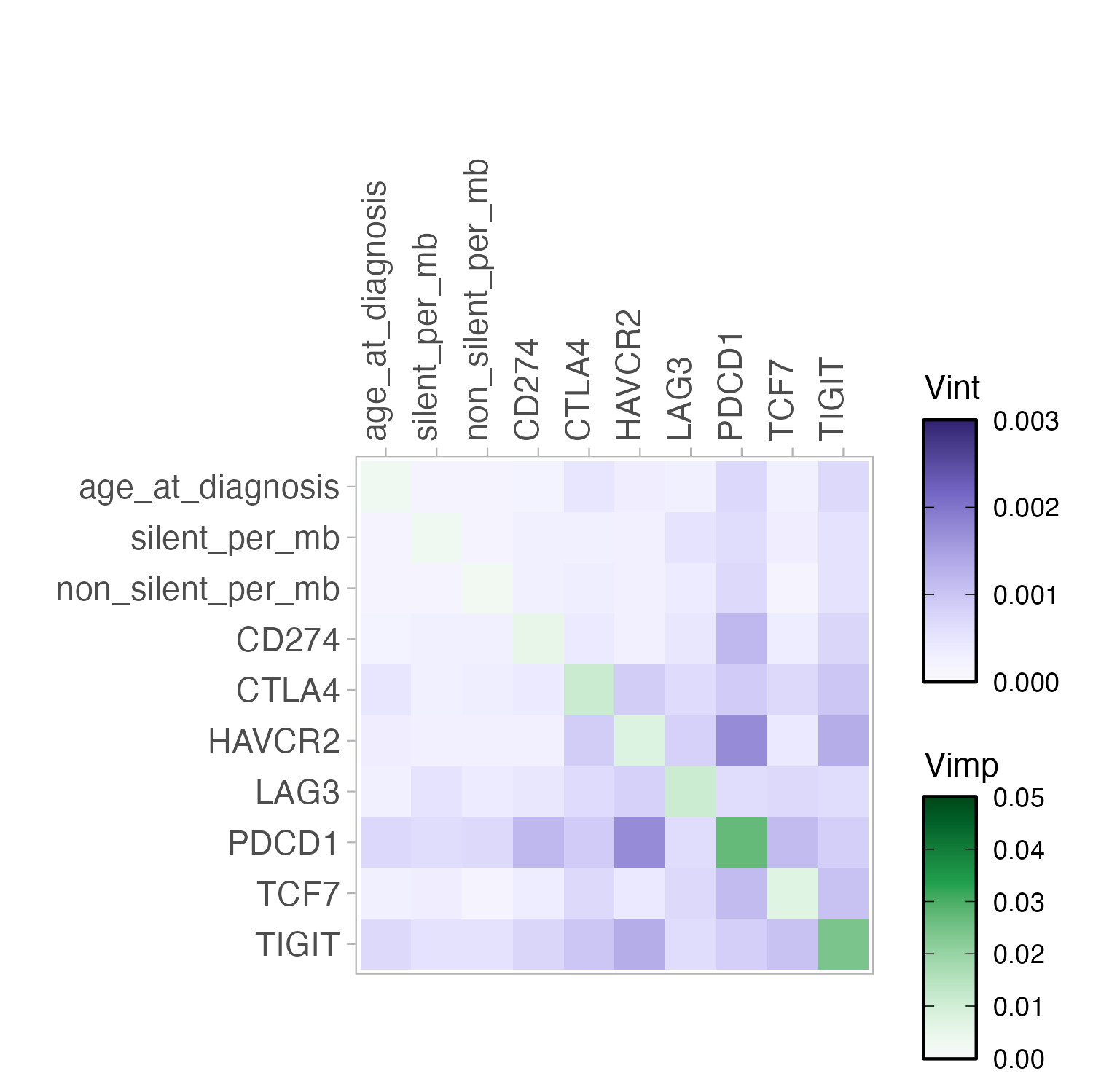}}  
    \caption{Predicted variable importance (Vimp) and variable interaction (Vint) plots for the two holdout sarcoma types. These were calculated by taking a weighted sum of VIVI plots for the remaining 5 sarcoma types.}
    \label{fig:vivi_pred}
\end{figure}

In \textbf{Figure \ref{fig:vivi_pred}}, we show the VIVI plot for the  subtypes included only in the test data by applying the weights of our method to the VIVI plots themselves. The fact that we can provide group-specific feature importances for groups not in the training data is one of the attractive features of our approach.


\begin{table}[!ht]
    \centering
    \caption{Mean squared error of a linear mixed model, decision tree, random forest, and our proposed weighted sum-of-trees method. The random forest method was performed twice, once with 5 trees and once with 100 trees. Our sum-of-trees method was also carried out twice. Sum of DTs is a weighted combination of 5 decision trees. Sum of RFs is a weighted combination of 5 random forests, each with 20 trees (100 trees total). The scale of the response variable (T-cell abundance) is standardized. The best performances are \textbf{bolded}. Key: LMM = linear mixed model, DT = decision tree, RF = random forest, SS = synovial sarcoma, MPNST = malignant peripheral nerve sheath tumor}
    \begin{tabular}{|c|c||c||c c||c c|}
    \hline
                & LMM & DT   & RF (5 trees)  & Sum of DTs & RF (100 trees) & Sum of RFs \\
         \hline
         SS \& MPNST (N=15)  & 2.964 & 1.786  & 0.826 & 0.643 & 0.694 & \bf{0.582}  \\
         \hline 
         SS (N=10)    & 4.261 & 2.238  & 1.084 & 0.721 & 0.924 &  \textbf{0.675} \\
         MPNST (N=5)  & 0.371 &  0.883 & 0.309 & 0.525 & \textbf{0.236} &  0.362 \\
         \hline
    \end{tabular}
    \label{tab:real_data_performance}
\end{table}

\begin{table}[!ht]
    \centering
    \caption{Weights learned during the first stage of the model. Key: DDLPS = dedifferentiated liposarcoma, MFS = myxofibrosarcoma, STLMS = soft tissue leiomyosarcoma, ULMS = uterine leiomyosarcoma, UPS = undifferentiated pleomorphic sarcoma, SS = synovial sarcoma, MPNST = malignant peripheral nerve sheath tumor}
    \begin{tabular}{|c|c|c|c|c|c|}
    \hline
                & DDLPS & MFS & STLMS & ULMS & UPS \\
         \hline
         SS    & 0.099& 0.006& 0.384& 0.4704& 0.040\\
         MPNST & 0.282& 0.012& 0.273& 0.185& 0.247 \\
         \hline
    \end{tabular}
    \label{tab:learned_weights}
\end{table}


\section{Discussion} \label{discussion}

When conducting our simulation tests, we experimented with using different learners to predict the group membership. We found that logistic regression and na{\"i}ve Bayes generally perform well. Random forest and decision tree classifiers can also perform well, but they sometimes yield worse results, depending on the simulation design parameters. We recommend logistic regression because it is easy to interpret and because the model coefficients can be inspected to identify important variables.

Changing the amount of global noise tended to degrade the performance of all methods while preserving their relative positions.
Our proposed weighted sum-of-trees method seems to perform best relative to other methods when the number of observations per group exceeds the number of groups, $n>K$. This makes sense because as $K$ is increased, the output probabilities from logistic regression (or whichever method you choose), will tend to be smaller and spread across a larger number of base learners, making the output prediction more like a random forest but without the bootstrap.

Our method does better as the magnitude of the random effects grows and $\sigma^2_\alpha$ gets larger. For the same reason, it also makes sense that LMM's performance remains steady as the random noise increases. 

We also see that incorporating random forests instead of decision trees into the second stage of the model does not yield a significant improvement in performance. The weighted sum-of-trees approach stands out for its strong performance while still retaining interpretability and competitive run times. Unlike random forests, which exhibit comparable performance, our method does not use bootstrapping of the observations and does not randomly sample features. For this reason, the base learners used in our method retain their interpretability. Furthermore, with one learner trained per in-sample group, the base models can be inspected, improving the potential for trust in clinical applications.

While we tested only data with continuous outputs, this method is easy to use with categorical outputs as well, without the need for a logit or probit link function in the second-stage models.


In the real data results, we see that our method performs better when compared to random forests with an equal number of trees. This is even more impressive considering that our method does not randomly sample features or bootstrap the data. This finding opens the door into how these tricks can be implemented into our method for even better performance.

We also see interesting results in the VIVI plots for the different types of sarcoma. Most significantly, MFS shows that \textit{PD-1} (\textit{PDCD1}) is a highly important  feature in MFS, with all other variables and interactions having weak importance. In STLMS and ULMS, on the other hand, \textit{TIGIT} is also important as a primary variable, and the interaction of \textit{PD-1} and \textit{TIM-3} (\textit{HAVCR2}) in STLMS is notable. This is a potentially very interesting translational finding: currently, tumor \textit{PD-L1} expression is a well-established clinical biomarker of immunotherapy response \citep{li2022biomarkers}. Our findings suggest that this marker may work well for MFS, as well as that additional genes and their interactions might offer an improved biomarker for other subtypes.

In our case study, we illustrated the utility of our method in predicting T-cell abundances in sarcoma from clinical and genomic variables. Models intended for clinical risk prediction should provide an understanding uncertainty alongside the point prediction for a specific patient \citep{riley2025uncertainty}. Various approaches have been explored for constructing prediction intervals for random forests, including parametric methods, quantile random forests, and conformal inference \citep{roy2020prediction}. \citet{roy2020prediction} recommend the jackknife version of the conformal inference method proposed by \citet{lei2018distribution}. As this method is agnostic to the prediction method being used, it could be applied to obtain prediction intervals for our weighted sum-of-trees approach.


Our proposed approach has some limitations.
One key consideration is that the weighted sum-of-trees model performs more competitively when the number of observations per group, $n$, is sufficiently large.
Another consideration in applying our proposed approach is that the group assignments are assumed to be known. In cases where the groups are not known a priori, group labels could be learned using unsupervised clustering prior to applying our approach. 
Finally, if the goal is to obtain predictions for test observations from both in-sample and out-of-sample groups, it might make sense to rely on a linear mixed model or mixed effects machine learning model for the in-sample groups, where it is possible to estimate the random effects, and apply our proposed approach primarily for the out-of-sample groups which were not encountered in the training data.

\section{Conclusions} \label{conclusions}

We find that our sum-of-trees method provides reliable performance over a range of simulation settings. 
The sum-of-trees method also becomes competitive with random forests, even though the sum-of-trees method does not use any bootstrapping of observations or random sampling of features.
Overall, our work shows that constructing decision trees and forests can be improved when clustered or grouped observations are present without the need for rigid model constraints.
In the future, we plan on investigating how this method translates to categorical and survival outputs. 





\newpage
\bibliographystyle{unsrtnat}
\bibliography{main}

\newpage
\section*{Additional tables and figures}\label{secA1}


\begin{table}[!ht]
    \centering
    {
    \caption{Average run times in seconds. Recall that $n$ is the number of observations per group, and thus the total number of observations used to train each method is equal to $n$ multiplied by $J$, which in this case is 16 (J=$0.8\times K=16$).}
    \begin{tabular}{|c c c c c|}
    \toprule
$n$ & 20 & 50 & 100 & 500 \\
\midrule
Linear Mixed Model &        2.346 &	2.790 &	2.898 &	5.601 \\
Decision Tree &	            0.004 &	0.008 &	0.017 &	0.100 \\
Random Forest &	            0.030 &	0.073 &	0.156 &	0.981 \\
BART &	                    0.188 & 0.392 &	0.807 &	4.921 \\
Weighted Sum-of-Trees &	    0.033 &	0.038 &	0.048 &	0.147 \\
Weighted Sum-of-Forests &   0.154 &	0.171 &	0.213 &	0.668 \\
         \bottomrule
    \end{tabular} \label{tab:runtime_data}
    }
\end{table}





\end{document}